\documentclass[aps,pre,twocolumn,superscriptaddress,showpacs]{revtex4}
\usepackage{amsmath,epsfig,amsfonts,amssymb,latexsym,times}
\begin{document}

{\small Phys. Rev. E \textbf{73}, 065106 (2006)}
  
\title{Synchronization is optimal in non-diagonalizable networks}

\author{Takashi Nishikawa}
\affiliation{Department of Mathematics, Southern Methodist University, Dallas, TX 75275}

\author{Adilson E. Motter}
\affiliation{Department of Physics and Astronomy, Northwestern University, Evanston, IL 60208}
\affiliation{CNLS and Theoretical Division, Los Alamos National Laboratory, Los Alamos, NM 87545}

\date{June 30, 2006}

\begin{abstract}
We consider the problem of maximizing the synchronizability of oscillator networks
by assigning weights and directions to the links of a given interaction topology.
We first extend the well-known master stability formalism to the case of non-diagonalizable 
networks.
We then show that, unless some oscillator is connected to all the others,
networks of maximum synchronizability are necessarily non-diagonalizable and can always be
obtained by imposing unidirectional information flow with normalized input strengths.
The extension makes the formalism applicable to all possible network structures,
while the maximization results provide insights into hierarchical structures
observed in complex networks in which synchronization plays a significant role.
\end{abstract}


\maketitle

Under extensive study in recent years is how the collective dynamics of a complex
network is influenced by the structural properties of the network \cite{recent},
such as clustering coefficient~\cite{wattsstrogatz1998}, average network distance~\cite{watts1999},
connectivity
distribution~\cite{barabasi1999},  assortativity~\cite{Newman2002}, and weight distribution
\cite{Motter2005b,Motter2005,Zhou2006}. The effects of these properties on synchronization has particularly
attracted the attention of researchers, partly because of the elegant analysis due to
Pecora and Carroll~\cite{pecora1998} which allows us to isolate the 
contribution
of
the network structure in terms of the 
eigenvalues of the coupling matrix.

Synchronizability of complex networks of oscillators generally has been shown to improve as the
average network distance decreases, with one notable exception:
in random scale-free networks,
which are characterized by a
strong heterogeneity of the 
connectivity
distribution \cite{barabasi1999}, 
synchronization was shown to become
more difficult as the heterogeneity increases~\cite{Nishikawa2003}, even though the average
network distance decreases at the same time. Motivated by this counter-intuitive effect,
researchers have pursued
ways to enhance the synchronizability of scale-free networks by introducing directionality and
weight to each link in the network~\cite{Motter2005b,Motter2005,Chavez2005}. 
A natural question arising in this context is: \textit{Given a network of oscillators with a fixed
topology of interactions, which assignment of weights and directions maximizes
its synchronizability?}  By maximization, we mean that the synchronized states are stable for the widest
possible range of the parameter representing the overall coupling strength.

The study of such a question not only provides us with insights into the dynamics of 
real-world complex networks but also guides us in designing large artificial networks.
Metabolic networks---the system of hundreds of interconnected biochemical reactions
responsible for the biomass and energy production in a cell---is a prototypic
example where the weights and directions of feasible links (metabolic fluxes) are adjusted 
to optimize fitness, which is likely to account for robustness of synchronized behavior
against environmental changes \cite{fishcer:2005}. Other 
examples 
range
from the enhancement of neuronal synchronization for a given topology of synaptic
connections in the brain, to the design of interaction schemes that optimize the performance
of computational tasks based on the synchronization of processes in computer networks 
\cite{korniss:2003}. The adjustment of flows in power grids and communication patterns
in social organizations are additional examples where directional and weighted patterns
can be favored because they can better facilitate the synchronized or coordinated behavior
on which the functioning of these networks is based.


Here we show that the answer to the question of maximum synchronizability 
falls outside the framework of the Pecora-Carroll analysis, which is built
on the assumption that the 
network dynamics can be linearly decomposed into eigenmodes, i.e., the
coupling matrix of the network is diagonalizable. Indeed, we show that maximally
synchronizable networks are always \emph{non-diagonalizable}
(except for the 
extreme configurations where a node is connected to all the others)
and can be constructed for any given interaction
topology by imposing that the network: ({\it i}) embeds an {\em oriented spanning tree}, 
({\it ii}) has {\em no directed loops}, and ({\it iii}) has {\em normalized} input strength in each node.
The fact that the 
networks
are not necessarily
diagonalizable has been largely overlooked in the literature, apparently because 
most previous works have focused 
on networks of symmetrically coupled oscillators,
which are guaranteed to be diagonalizable. However, 
the same does not hold true in general when the network is directed,
as required in the realistic modeling of many complex systems.
Here we develop a new theory that extends the Pecora-Carroll
analysis to the case of non-diagonalizable networks.  
We show that in this case the
synchronizability  is still determined 
by the eigenvalues of the 
coupling matrix, but the speed at which the system converges toward the synchronized
state may be significantly slower. 
This theory is a first example of going beyond the traditional framework for studying complex systems based on either decomposition into eigenmodes or some sort of superposition principle.

Consider $n$ identical oscillators whose individual dynamics without coupling is 
governed by $\dot{\mathbf{x}} = \mathbf{F}(\mathbf{x})$, $\mathbf{x}\in I\!\! R^m$. 
Now consider the network of these oscillators coupled via an output signal function 
$\mathbf{H}: I\!\! R^m \to I\!\! R^m$ along a network with
a symmetric adjacency matrix $A=(A_{ij})$ defined by $A_{ij} = 1$ if oscillators
$i$ and $j$ ($\neq i$) are connected and $A_{ij} = 0$ otherwise.  
Let $W_{ij}\ge 0$ denote the strength of the coupling that oscillator $i$
receives from $j$.  Thus, $A$ represents the topology of interactions
and $W=(W_{ij})$ represents the assignment of weights and directions.  The system of
equations governing the dynamics of the oscillator network  
can then be written as $\dot{\mathbf{x}}_i = \mathbf{F}(\mathbf{x}_i) + \sigma \sum_{j=1}^n A_{ij}W_{ij}[\mathbf{H}(\mathbf{x}_j) - \mathbf{H}(\mathbf{x}_i)]$
or, equivalently,
\begin{equation}
\dot{\mathbf{x}}_i =\mathbf{F}(\mathbf{x}_i) - \sigma \sum_{j=1}^n L_{ij}\mathbf{H}(\mathbf{x}_j),\quad i=1,\ldots ,n,
\label{eqn:main}
\end{equation}
where $\sigma$ is the parameter controlling the overall coupling strength
and $L = (L_{ij})$ is the 
coupling
matrix of the directed weighted network, defined
by $L_{ij} = -A_{ij}W_{ij}$ if $i \neq j$ and $L_{ii} = -\sum_{j\neq i} L_{ij}$. 
Note that $L$ is not necessarily symmetric because the network is not constrained to
be undirected.

The maximization problem considered in this paper can
be formulated as follows. For a given topology of interactions 
between oscillators (represented by $A$), we want to find the 
assignment of weights and directions (represented by $W$) that
maximizes the synchronizability of the network. In order to address
this question, 
we need a condition for the network to synchronize.
For any solution $\mathbf{x}=\mathbf{s}(t)$ of the individual dynamics
$\dot{\mathbf{x}} = \mathbf{F}(\mathbf{x})$, the completely synchronous
state $\mathbf{x}_i = \mathbf{s}(t)$, $i=1,\dotsc,n$ is automatically
a solution of the entire system~\eqref{eqn:main}.
The question then is  to determine when
this solution is stable against small perturbations.  This synchronization 
condition can be derived by extending the linear stability analysis of 
Pecora and Carroll~\cite{pecora1998} to the case where $L$ is not
necessarily diagonalizable, as follows.

The starting point of our analysis is the observation that, for any $n \times n$
matrix $L$, there exists an invertible matrix $P$ of generalized eigenvectors of $L$
which transforms $L$ into Jordan canonical form as $P^{-1}LP = J$, where
\begin{equation}
\label{eqn:jordan}
J = \begin{pmatrix}
0 & & &\\
& B_1 & &\\
& & \ddots & \\
& & & B_l
\end{pmatrix},\;\;
B_j = \begin{pmatrix}
\lambda & & & \\
1 & \lambda & & \\
& \ddots & \ddots & \\
& & 1 & \lambda \\
\end{pmatrix},
\end{equation}
and $\lambda$ is one of the (possibly complex) eigenvalues of $L$. 
The stability of the synchronous solution of Eq.~\eqref{eqn:main}
is determined by the variational equation
$ \dot{\xi} = D\mathbf{F}(\mathbf{s})\xi - \sigma D\mathbf{H}(\mathbf{s})\xi L^T$,
where $\xi = (\boldsymbol{\xi}_1, \ldots ,\boldsymbol{\xi}_n)$ and $\boldsymbol{\xi}_i$
is the perturbation to the $i$th oscillator.
By applying the change of variable
$\eta = \xi P^{-T}$, we get
\begin{equation}
\label{eqn:variational2}
\dot{{\eta}} = D\mathbf{F}(\mathbf{s}){\eta} - \sigma D\mathbf{H}(\mathbf{s}){\eta} J^T.
\end{equation}
Each block of the Jordan canonical form corresponds to a subset of equations in
\eqref{eqn:variational2}. For example, if block $B_j$ is $k \times k$, then it
takes the form
\begin{align}
\dot{\boldsymbol{\eta}}_{1} &= [D\mathbf{F}(\mathbf{s})
- \alpha D\mathbf{H}(\mathbf{s})]\boldsymbol{\eta}_{1} \label{eqn:eta1}\\
\dot{\boldsymbol{\eta}}_{2} &= [D\mathbf{F}(\mathbf{s}) 
- \alpha D\mathbf{H}(\mathbf{s})]\boldsymbol{\eta}_{2} - \sigma D\mathbf{H}(\mathbf{s})\boldsymbol{\eta}_{1} \label{eqn:eta2}\\
&\cdots\nonumber\\
\dot{\boldsymbol{\eta}}_{k} &= [D\mathbf{F}(\mathbf{s}) 
- \alpha D\mathbf{H}(\mathbf{s})]\boldsymbol{\eta}_{k} - \sigma D\mathbf{H}(\mathbf{s})\boldsymbol{\eta}_{k-1}, \label{eqn:eta3}
\end{align}
where $\alpha=\sigma\lambda$ and $\boldsymbol{\eta}_{1},\boldsymbol{\eta}_{2},\ldots,\boldsymbol{\eta}_{k}$ 
are perturbation modes in the generalized eigenspace of eigenvalue $\lambda$.  

For $\alpha$ regarded as a complex parameter, Eq.~\eqref{eqn:eta1} is 
a master stability equation
and its largest Lyapunov exponent $\Lambda(\alpha)$, called 
master stability function~\cite{pecora1998}, 
determines the stability of Eq.~\eqref{eqn:eta1}: 
it is linearly stable 
iff $\Lambda(\sigma\lambda) < 0$. 
The condition for Eq.~\eqref{eqn:eta2} to be stable is apparently more involved but
can be 
formulated
as follows. The linear stability of Eq.~\eqref{eqn:eta1}
implies that $\boldsymbol{\eta}_1$ converges to zero exponentially as $t \to \infty$.
Assuming that the norm of $D\mathbf{H}(\mathbf{s})$ is bounded,
we have that the second term in Eq.~\eqref{eqn:eta2} is exponentially small.
Then, the same condition $\Lambda(\sigma\lambda) < 0$, now applied to Eq.~\eqref{eqn:eta2}, 
guarantees 
the stabilizing effect of 
both the first and second terms, 
resulting in exponential convergence of $\boldsymbol{\eta}_2$ to zero as $t \to \infty$.
The same argument applied repeatedly shows that $\boldsymbol{\eta}_3,\ldots,\boldsymbol{\eta}_k$ must also converge to zero if
$\Lambda(\sigma\lambda) < 0$.  This shows
that $\Lambda(\sigma\lambda) < 0$ is a necessary and sufficient condition for the linear
stability of the equations corresponding to each full block $B_j$.
This condition is valid not only in diagonalizable~\cite{pecora1998} but also in
non-diagonalizable networks.

However, it is worthwhile noting a crucial difference between the diagonalizable and
non-diagonalizable cases.  If $L$ is diagonalizable, 
then all Jordan blocks are $1 \times 1$, so there would be no equations like
\eqref{eqn:eta2} or \eqref{eqn:eta3}, and each mode of perturbation 
is decoupled from the others. Thus, the exponential convergence occurs
independently and simultaneously.  
On the other hand, if $L$ is not diagonalizable, some modes
of perturbation may suffer from a long transient.  For instance,
if we have a network of linearly coupled phase oscillators, $\dot\theta_i=\omega-\sigma\sum_j L_{ij} \theta_j$, $\theta_i \in S^1$, 
then we can explicitly solve Eqs.~\eqref{eqn:eta1}--\eqref{eqn:eta3}  for the solution $s(t) = \omega t$ to obtain the last perturbation mode $\eta_k= e^{-\alpha t}\sum_{i=0}^{k-1} c_i t^i$, 
where the constants $c_i$ depend on the initial condition.
Therefore, the larger the size $k$ of the Jordan block, the longer the transient.

Turning our attention back to the maximization problem, we first note that
the eigenvalues $\lambda_1, \ldots ,\lambda_n$ of matrix $L$ can be
ordered such that $0=\lambda_1 \le\text{Re}\,\lambda_2 \le \ldots \le \text{Re}\,\lambda_n$,
where one eigenvalue is always zero because $L$ has zero row sum and all the others
are guaranteed to have nonnegative real parts because of the Gerschgorin Circle Theorem.
Thus, taking all the Jordan blocks into account,
it follows from our stability analysis that the synchronous solution is stable if and
only if 
\begin{equation}
\mbox{$\Lambda(\sigma\lambda_i) < 0$ for $i=2,\ldots , n$}.
\end{equation} 
$\Lambda(\sigma\lambda_1)=\Lambda(0)\ge 0$ is the largest Lyapunov exponent of the
individual oscillators and corresponds to the stability along the synchronization manifold.
We next
note that Re$\,\lambda_2 > 0$ if and only if the network embeds an
oriented spanning tree, i.e., there is a node from which all other nodes can be reached
by following
directed links. This condition follows from the recent Ref.~\cite{Wu2005}
and generalizes the notion of connecteness to directed networks. We assume this condition 
here to ensure that the network is compatible with 
synchronization.

In most of the previously studied cases, 
the master stability function $\Lambda(\alpha)$, determined by $\mathbf{F}$, $\mathbf{H}$, and $\mathbf{s}$, 
has been found to be negative in a single convex bounded
region of the complex plane~\cite{conjugate}.
This implies the existence of a single interval 
($\sigma_{\min}$, $\sigma_{\max}$)
of the overall coupling strength $\sigma$ for which synchronization is stable. Thus, the synchronizability
of the network can be measured in terms of the relative interval $\sigma_{\max}/\sigma_{\min}$:
the network becomes {\it more
synchronizable} as 
$\sigma_{\max}/\sigma_{\min}$ becomes larger.
In the special case of undirected networks,
the eigenvalues of $L$ are real, and this measure of synchronizability is proportional to the ratio 
$\lambda_2/ \lambda_n$
\cite{barahona2002}.

A critical observation is that in order for the ratio $\sigma_{\max} / \sigma_{\min}$ 
to achieve absolute maximum 
for any given $\Lambda(\alpha)$ with a convex stability region, 
all nonzero eigenvalues must be  real and equal to each other.
The condition that the eigenvalues must be real follows from
the convexity of the stability region and the fact that
complex eigenvalues appear in conjugate pairs, 
while the condition that they must be equal follows
from the fact that, for real eigenvalues, the  ratio $\sigma_{\max} / \sigma_{\min}$ 
is proportional to $\lambda_2/\lambda_n$.
Thus, a network 
with 
\begin{equation}
0=\lambda_1 < \lambda_2=\cdots=\lambda_n
\label{eq.8}
\end{equation} 
has the {\it widest possible range of coupling strength} in which synchronization is stable,
independently of the individual node dynamics $\mathbf{F}$, output function $\mathbf{H}$, and
synchronous state $\mathbf{s}$, as long as the stability region is convex \cite{non_convex}.

Under the mild
assumption that the interaction topology
allows no oscillator to interact with all the other oscillators, {\it any maximally 
synchronizable network is necessarily non-diagonalizable}. 
This
comes from the fact that if $L$ is diagonalizable and satisfies the optimality
condition (\ref{eq.8}) with nonzero eigenvalues equal to $\lambda>0$, then all
the rows of the characteristic matrix $L-\lambda I$ must be equal. In terms of
the network structure, this means that each node must either have uniform output
to all the other nodes
(at least one of them must do so)
or have no output at all.
These exceptional cases include globally connected networks and directed star
configurations.  
However, it is uncommon in a large complex
network that an oscillator can communicate with all the other oscillators.
Therefore, our extension of the master stability analysis to 
non-diagonalizable networks was indeed necessary to properly address the
optimization problem.



\begin{figure}[t]
\begin{center}
\epsfig{figure=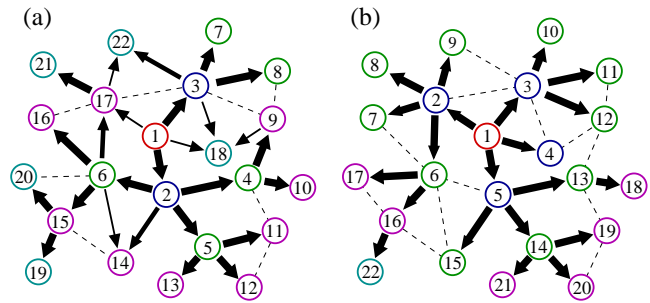,width=8.5cm}
\caption{(Color online) (a) Example of optimal assignment of weights and directions within a given interaction topology.  
The total input strength in each node is normalized to $\lambda$, where thick, medium, and thin arrows indicate weight $\lambda$,  $2\lambda/3$, and $\lambda/3$, respectively, and dashed lines have zero weight.   
Nodes are numbered and colored to show the hierarchical structure in which connections are only from a higher level to a lower level, with no feedback loops.
(b) Example of oriented spanning tree within the same 
interaction topology as in (a), constructed by the breadth-first search.  
}
\label{fig:ost}
\end{center}
\end{figure}

Having observed that optimal networks are rarely diagonalizable,
we now show that, for {\it any connected topology of interactions}, 
there are assignments of directions and weights for which {\it the
resulting network is non-diagonalizable and maximally synchronizable}.
We first note that maximum synchronizability can always be achieved 
by imposing that the network ({\it i}) embeds an oriented spanning tree,
({\it ii}) has no directed loops, and ({\it iii}) has normalized input strengths in
each node, i.e., the total input is the same for all nodes that receive input.
Condition ({\it i}) guarantees that  Re$\,\lambda_2 > 0$,  condition ({\it ii}) 
guarantees that the eigenvalues are real, and condition ({\it iii}) then
implies the identity (\ref{eq.8}) among the nonzero eigenvalues.
In such optimal networks, we can always rank the nodes so that each 
node receives inputs only from nodes that are higher in the ranking 
(see Fig.~\ref{fig:ost}(a) for an example).  
In this hierarchical 
structure, information flows only from top to bottom 
of the ranking, without feedback.
The optimality can be formally confirmed by noting that indexing 
nodes  according to the ranking makes $L$ a lower triangular matrix 
with  $0,\lambda,\ldots,\lambda$ on the diagonal, which means that 
$\lambda_2=\cdots=\lambda_n=\lambda$, where $\lambda>0$ is the total
input strength in $n-1$ of the nodes. An important class of such 
maximally synchronizable networks consists of the oriented spanning 
trees themselves, where the normalization condition leads to 
uniform weights for all links of the tree (see Fig.~\ref{fig:ost}(b)
for an example).  This example shows that any interaction topology
admits at least $n-1$, but usually many more, optimal non-diagonalizable 
networks. Indeed, from the Matrix-Tree Theorem it follows that the number 
of all oriented spanning trees is $\Pi_{i=2}^{n}\mu_i$, where $\mu_2,\ldots,\mu_n$ 
are the nonzero 
Laplacian eigenvalues of the underlying {\it undirected} network defined 
by matrix $A$. For a globally connected network, for example, the number 
is $n^{n-1}$, which is huge even for
relatively small networks. 
All these oriented spanning trees are non-diagonalizable, except for the star configuration. 
Oriented spanning trees can be explicitly constructed by 
the well-known procedure called the breadth-first search, which spans
all nodes starting from an arbitrary root node.

Physically, the optimality conditions ({\it i})-({\it iii}) can be understood as follows.  
The top node in the ranking receives no input and acts as a {\it master oscillator} that 
dominates  the network dynamics. If the coupling strength $\sigma$ is chosen so that 
$\Lambda(\sigma\lambda) < 0$, 
then the oscillators that are
immediately lower in the hierarchy and receive input from
the master will synchronize themselves with the master.   Any
oscillator receiving input only from these oscillators and the master
must also synchronize, since normalization of the total input
strength makes the equation effectively look as if it were receiving
input from a single oscillator that is synchronized with the master.  Repeating the same argument 
for the rest of the network, we see that under conditions ({\it i})-({\it iii})
all oscillators must eventually synchronize and they do so for
the entire range of $\sigma$ 
where $\Lambda(\sigma\lambda) < 0$.

Interestingly, {\it undirected} tree 
networks have
been found to be among the most difficult to 
synchronize~\cite{Yook2005}, in striking contrast to our result that {\em directed} spanning trees 
lead to the most synchronizable
configurations.  
  This highlights the significance 
of directionality of the interactions in determining the synchronizability of the networks~\cite{Zheng2000}.  On the other hand, the choice of the master oscillator
in a maximally synchronizable network is completely arbitrary, 
despite the intuition that the nodes with largest 
connectivity would be the most natural choice.
Moreover, 
the directions of the links
in such a network 
are not necessarily related to the properties of the nodes they connect,
even though there has been a suggestion that it would be related to the age of the nodes~\cite{Hwang2005}.
In contrast, under the stricter constraint 
that all feasible input connections
have the same strength in each node,
it was found~\cite{Motter2005b,Motter2005} that maximum synchronizability 
is achieved when 
the individual
input strength is inversely proportional to the 
connectivity of the node, 
which is consistent with our result that normalization is key to ensuring optimality.

The optimality conditions ({\it i})-({\it iii}) suggest that in designing a network for which synchronization is desired, it is generally advantageous
to avoid feedback loops and to normalize input strength.
Because these conditions typically lead to assigning nonzero weights only to a subset of all 
possible links,
this interesting result
can be interpreted as a synchronization version of the paradox of Braess for traffic flow~\cite{braess},
in which removing links leads counter-intuitively to improved performance of the network.
Furthermore, such assignment of weights not only maximize the synchronizability, but also {\it minimize} the coupling cost.
The coupling cost can be defined as the sum of the input strengths
of all nodes at the synchronization threshold~\cite{Motter2005b}.  
If $\Lambda(\alpha) < 0 $ in $(\alpha_1, \alpha_2)$, then the coupling cost for any network can only be as small as $\alpha_1(n-1)$, which can be achieved by networks with global uniform coupling.  A surprising fact, however, is that this minimum can also be achieved by the maximally synchronizable networks as well.
In other words, our optimality conditions allow a network constrained
by an arbitrary topology to synchronize with the best possible efficiency.
It is interesting to point out that efficiency optimization of traffic flow on a transportation network model leads to a hierarchical structure similar to that possessed by our maximally synchronizable networks~\cite{Barthelemy2006}.

Our characterization of the maximally synchronizable networks can be used to test the 
widely assumed hypothesis that synchronizability plays an important role in the evolution of many real-world complex networks. The loop structure of
the metabolic network of {\em E. coli} suggests that having fewer loops may have been beneficial for the cell (the details will be published elsewhere), while recent experimental findings~\cite{neuron} suggest the significance of hierarchical structures in neuronal networks.  Exploring more real data to systematically test this hypothesis is of critical fundamental for a better understanding of complex networks.

A. E. M. was supported by DOE under Contract No. W-7405-ENG-36.


\begin{thebibliography}{99}
\bibitem{conjugate}
In addition, it follows from the complex conjugate of Eq.~\eqref{eqn:eta1}
that $\Lambda(\alpha)$ is always symmetric about the real axis.
\bibitem{non_convex}
If the number $n$ of oscillators is even, the same would hold true even
for non-convex regions.
\end{thebibliography}
\end{document}